\documentclass[aps,twocolumn,groupedaddress]{revtex4}

\usepackage{graphics}
\usepackage{graphicx}

\def\fun#1#2{\lower3.6pt\vbox{\baselineskip0pt\lineskip.9pt
\ialign{$\mathsurround=0pt#1\hfil##\hfil$\crcr#2\crcr\sim\crcr}}}

\begin{document}

\title{QED Radiative Corrections in Processes of Exclusive Pion
Electroproduction}

\date{\today}
\author{A.~Afanasev$^{a}$, I.~Akushevich$^{b}$, V.~Burkert$^{a}$, K.~Joo$^{a}$}
\affiliation{
a) Jefferson Lab, Newport News, VA 23606\\
b) Duke University, Durham, NC 27708
}

\begin{abstract}
Formalism for radiative correction (RC) calculation in exclusive pion
electroproduction on the proton is presented. A FORTRAN code EXCLURAD is developed for the RC
procedure. The numerical analysis is done in the kinematics of current Jefferson Lab
experiments.
\end{abstract}

\maketitle

\section{Introduction}
Understanding the electromagnetic transition
amplitudes from the nucleon ground state to excited states provides
valuable insights into the electromagnetic structure of the nucleon. 
Exclusive pion electroproduction is one of the major sources to provide
the most direct information about the spatial and spin structure of
the excited states. 
With the development of high intensity and high
duty--factor electron beam with high degree of polarization, this field reaches 
a new level of quality. For the past several years, exclusive pion
electroproduction has been the main subject of extensive studies at 
various accelerator laboratories such as
MIT-Bates, ELSA, MAMI and Jefferson Lab. 

New measurements with the CLAS detector at Jefferson Lab/Hall B \cite{nstar}
are expected
to greatly improve the systematic and statistical precision and cover
a wide kinematic range in four--momentum transfer $Q^2$ and invariant
mass $W$, as well as the full angular range of the resonance decay
into the nucleon--pion final state:
\begin{eqnarray} \label{process}
 e(k_1) + p (p) &\longrightarrow & e'(k_2) + \pi^+(p_h) + n(p_u),
\nonumber \\
 e(k_1) + p (p) &\longrightarrow & e'(k_2) + p(p_h) + \pi^0(p_u),
\end{eqnarray}
where $p_h (p_u)$ denotes the momentum of the detected (undetected) hadron.  
Adequate calculation of radiative corrections (RC) becomes 
important in interpreting the measured observables such as unpolarized
coincidence cross sections and polarization asymmetries.
 
While solving the RC problem, people are commonly referred to the classical
approach developed by Mo and Tsai \cite{MoTsai} and used for inclusive and
elastic electron scattering for decades. However, this approach cannot be 
directly applied for exclusive pion electroproduction due to the following 
 reasons.

First, we now deal with {\it exclusive} electroproduction,
where the hadron is detected in addition to the final electron. It
reduces the room for phase space allowed for the final radiated
photon. The formulas of Mo and Tsai as well as any other inclusive formulas 
cannot be applied for this case without additional strong assumptions. 

Second, there are only contributions of two structure functions in the inclusive case. The exact
formalism  for the  exclusive process requires consideration of four
structure functions for the unpolarized case with additional angular dependence associated
with them. Note that transition to  the case of two 
structure functions would not be possible even within realistic
approximations. Mo and Tsai's approach predicts neither RC to polarization
asymmetries, nor dependencies on the outgoing hadron angles.

\begin{figure}[t]
\begin{center}
\begin{tabular}{ccc}
\begin{picture}(70,100)
\put(30,60){\line(2,-1){20.}}
\put(50,50){\line(2,1){20.}}
\put(50,17.5){\circle*{10.}}
\multiput(50,28)(0,8){3}{\oval(4.0,4.0)[r]}
\multiput(50,24)(0,8){4}{\oval(4.0,4.0)[l]}
\put(30,10){\line(2,1){15.}}
\put(30,9){\line(2,1){15.}}
\put(55,18.5){\line(2,1){15.}}
\put(55,17.5){\line(2,-1){15.}}
\put(55,16.5){\line(2,-1){15.}}
\put(50,-10){\makebox(0,0){\small a)}}
\put(20,65){\makebox(0,0){\small $e(k_1)$}}
\put(80,65){\makebox(0,0){\small $e'(k_2)$}}
\put(20,8){\makebox(0,0){\small $p(p)$}}
\put(80,8){\makebox(0,0){\small $p_u$}}
\put(80,30){\makebox(0,0){\small $p_h$}}
\put(37,39){\makebox(0,0){\small $\gamma^*(q)$}}
\end{picture}
&
\begin{picture}(70,100)
\multiput(40,57)(0,8){3}{\oval(4.0,4.0)[r]}
\multiput(40,61)(0,8){3}{\oval(4.0,4.0)[l]}
\put(30,60){\line(2,-1){20.}}
\put(50,50){\line(2,1){20.}}
\put(50,17.5){\circle*{10.}}
\multiput(50,28)(0,8){3}{\oval(4.0,4.0)[r]}
\multiput(50,24)(0,8){4}{\oval(4.0,4.0)[l]}
\put(30,10){\line(2,1){15.}}
\put(30,9){\line(2,1){15.}}
\put(55,18.5){\line(2,1){15.}}
\put(55,17.5){\line(2,-1){15.}}
\put(55,16.5){\line(2,-1){15.}}
\put(50,-10){\makebox(0,0){\small b)}}
\put(52,70){\makebox(0,0){\small $\gamma (k)$}}
\end{picture}
&
\begin{picture}(70,100)
\multiput(60,57)(0,8){3}{\oval(4.0,4.0)[r]}
\multiput(60,61)(0,8){3}{\oval(4.0,4.0)[l]}
\put(30,60){\line(2,-1){20.}}
\put(50,50){\line(2,1){20.}}
\put(50,17.5){\circle*{10.}}
\multiput(50,28)(0,8){3}{\oval(4.0,4.0)[r]}
\multiput(50,24)(0,8){4}{\oval(4.0,4.0)[l]}
\put(30,10){\line(2,1){15.}}
\put(30,9){\line(2,1){15.}}
\put(55,18.5){\line(2,1){15.}}
\put(55,17.5){\line(2,-1){15.}}
\put(55,16.5){\line(2,-1){15.}}
\put(50,-10){\makebox(0,0){\small c)}}
\end{picture}
\end{tabular}
\begin{tabular}{rr}
\begin{picture}(70,100)
\put(30,60){\line(2,-1){20.}}
\put(50,50){\line(2,1){20.}}
\put(50,17.5){\circle*{10.}}
\multiput(50,28)(0,8){3}{\oval(4.0,4.0)[r]}
\multiput(50,24)(0,8){4}{\oval(4.0,4.0)[l]}
\multiput(42,55)(4,4){3}{\oval(4.0,4.0)[lt]}
\multiput(42,59)(4,4){2}{\oval(4.0,4.0)[br]}
\multiput(50,63)(4,-4){3}{\oval(4.0,4.0)[tr]}
\multiput(54,63)(4,-4){2}{\oval(4.0,4.0)[bl]}
\put(30,10){\line(2,1){15.}}
\put(30,9){\line(2,1){15.}}
\put(55,18.5){\line(2,1){15.}}
\put(55,17.5){\line(2,-1){15.}}
\put(55,16.5){\line(2,-1){15.}}
\put(50,-10){\makebox(0,0){\small d)}}
\end{picture}
&
\begin{picture}(70,100)
\put(30,60){\line(2,-1){20.}}
\put(50,50){\line(2,1){20.}}
\put(50,17.5){\circle*{10.}}
\multiput(50,48)(0,8){1}{\oval(4.0,4.0)[r]}
\multiput(50,44)(0,8){1}{\oval(4.0,4.0)[l]}
\multiput(50,28)(0,8){1}{\oval(4.0,4.0)[r]}
\multiput(50,24)(0,8){1}{\oval(4.0,4.0)[l]}
\put(50,36){\circle{12.}}
\put(30,10){\line(2,1){15.}}
\put(30,9){\line(2,1){15.}}
\put(55,18.5){\line(2,1){15.}}
\put(55,17.5){\line(2,-1){15.}}
\put(55,16.5){\line(2,-1){15.}}
\put(50,-10){\makebox(0,0){\small e)}}
\end{picture}
\end{tabular}
\end{center}
\vspace{0.5cm}
\caption{
\protect
Feynman diagrams contributing to the Born and the next--order
electroproduction cross sections. a) Born process, b) and c) Bremsstrahlung,
d) Vertex correction, and e) Vacuum polarization. The momentum $p_h(p_u)$ is assigned
to the detected (undetected) hadron.
}
\label{feyn}
\end{figure}
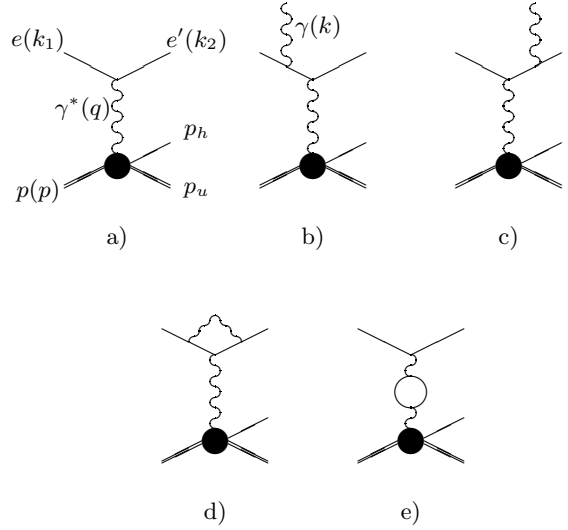

The third reason is a known shortcoming of the Mo and
Tsai's approach, namely, the dependence on an unphysical parameter splitting soft
and hard regions of the phase space of radiated photon in order to cancel the
infrared divergence. 

In our approach, which is based on  a covariant procedure of infrared divergence 
cancellation proposed
by Bardin and Shumeiko in Ref.\cite{BSh}, such an  unphysical parameter is not required.
Previously, this approach was applied for the calculation of RC for inclusive
\cite{ABSh,KSh1,ASh}, semi--inclusive \cite{Soroko,AkuSor} and
exclusive diffractive \cite{Ak} reactions. Recent reviews of the approach,
higher order effects and calculation for specific experiments can
be found in papers \cite{ABK,AKS,AISh2001}. Based on these results, a Fortran
code POLRAD \cite{POLRAD20} for RC calculation in polarized inclusive and
semi--inclusive processes was developed. Besides, some specific tasks such
as Monte Carlo generator RADGEN \cite{RADGEN}, MC approach to
diffractive
vector meson electroproduction \cite{DIFF2}, RC to spin--density
matrix elements in exclusive vector meson production \cite{AKuzh},
quasielastic tail for polarized He--3 target \cite{ARST}
were solved. 
Recently Bardin--Shumeiko approach was
applied to the measurements of elastic polarized  electron--proton
scattering at Jefferson Lab \cite{pep1,pep2}. 
A comprehensive analysis of results
obtained in Ref. \cite{MoTsai} and \cite{BSh}
was made in Refs. \cite{RADGEN, KuzhSh}.

Note that there are also other approaches for calculation of RC in
electroproduction processes. For example, the results of Ref.
\cite{spiesberger} are actively used for HERA experiments, while the
approach developed in papers \cite{Kuraev} is applied to specific
measurements at Jefferson Lab \cite{AAMsf}.

The Feynman diagrams needed to calculate RC are presented
in Fig.\ref{feyn}. They include QED processes of radiation of an unobserved real
photon, vacuum polarization and lepton--photon vertex corrections. These processes
give the largest contribution due to a large logarithmic term $\ln(Q^2/m_e^2)$. They can be
calculated exactly from QED, and uncertainties of such a calculation are only due to
the fits and data used for the hadronic structure functions. These uncertainties are
demonstrated in the present article. Additional mechanisms (box--type
diagrams, emission by hadrons) are smaller by about an order of magnitude  
and they contain considerable theoretical uncertainties. Most recent studies of two--photon 
exchange effects for elastic $ep$--scattering were reported in Refs.\cite{MaxTj,AAExcl02}. 
Previous RC calculations \cite{AISh95} for inclusive polarized deep--inelastic scattering 
included above--mentioned additional mechanisms. Generalization to the case of meson
electroproduction will be subject to a separate study.

The paper is organized as follows. We introduce kinematics and definitions (Section II),
derive a cross section of the radiative process (Section III), 
solve the problem of infrared divergence (Section IV), obtain RC in the leading log approximation
(Section V), verify the relation between exclusive and inclusive RC (Section VI), perform numerical
analysis (Section VII), and summarize the results in Section VIII.

\begin{figure}
\includegraphics[height=8cm,width=6cm]{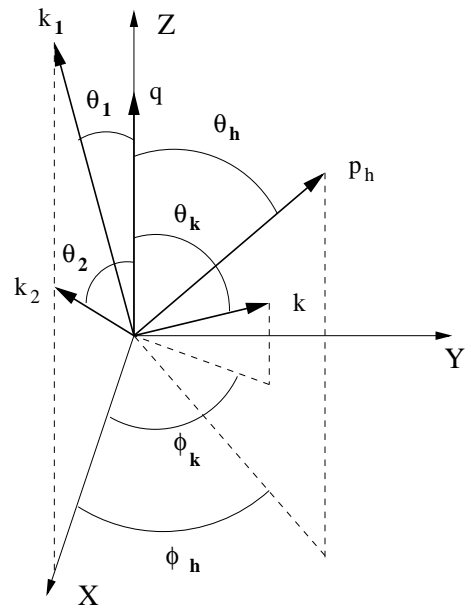}
\caption{Definition of momenta and angles in center-of-mass frame.}
\label{frame}
\end{figure}

\section{Kinematics and Born process}

At the Born level (Fig.\ref{feyn}a), the cross section of the processes 
(\ref{process}) is described by four kinematic variables. Following tradition, we
choose them as squared virtual photon momentum $Q^2$, invariant mass of
initial proton and the virtual photon $W$, 
and detected pion (or proton) angles $\theta_h$ and
$\phi_h$ in the center--of--mass of the final hadrons. The obtained
formulas are equally valid for both electron and muon scattering.

We use the following Lorentz invariants defined from leptonic 4--momenta:
\begin{eqnarray}\label{lepkin}
&&S=2k_1p, \; X=2k_2p, \; Q^2=-(k_1-k_2)^2,\;
\nonumber\\ &&
u_1=S-Q^2,\; u_2=X+Q^2,\; \lambda_{1,2}=u_{1,2}^2-4m^2W^2,
\nonumber\\ &&
S_{p,x}=S\pm X, \; \lambda_S=S^2-4m^2M^2
\nonumber\\ &&
W^2=S_x-Q^2+M^2, \lambda_q=S_x^2+4M^2Q^2,
\nonumber\\ &&
\lambda=Q^2u_1u_2-Q^4W^2-m^2\lambda_q,
\end{eqnarray}
where $m(M)$ is the lepton (proton) mass.

We use the c.m. system of virtual photon ($q=k_1-k_2$) and
initial nucleon. The axis $OZ$ is chosen
along $\vec q$ (see Fig. \ref{frame}).
The energies and angles (Fig.\ref{frame}) 
in the selected frame can be expressed in terms of  invariants (\ref{lepkin})
as follows:
\begin{eqnarray}
&&E_{1,2}={u_{1,2}\over 2W},\;
E_{q}={S_x-2Q^2\over 2W},\;
E_{p}={S_x+2M^2\over 2W},\;
\nonumber\\ &&
p_{1,2}={\sqrt{\lambda_{1,2}}\over 2W},\;\;\;
p_{p}=p_q={\sqrt{\lambda_q}\over 2W},
\nonumber\\ &&
\cos\theta_{1,2}={u_{1,2}(S_x-2Q^2)\pm 2Q^2W^2\over
\sqrt{\lambda_{1,2}}\sqrt{\lambda_q}},
\nonumber\\ &&
\sin\theta_{1,2}={2W\sqrt{\lambda}\over
\sqrt{\lambda_{1,2}}\sqrt{\lambda_q}}.
\end{eqnarray}
Here subscripts $1,2,q,p$ denote  the initial lepton, final lepton, virtual
photon and initial proton, respectively.

All the introduced kinematic variables have the same definition for both  Born
and radiative kinematics. But the situation is different for the final hadrons.
 The reason is that the energy of the observed hadron is not fixed
by  measurements of the chosen kinematic variables: $Q^2$, $W^2$,
$\theta_{h}$
$\phi_{h}$. As a result, it is different for these cases. In the Born case
the hadron energy can be defined via conservation laws, but for the radiative
process it depends on the unknown energy of the radiated photon. In order to
distinguish between the variables, we use the superscript '0' for the quantities
calculated for Born kinematics:
\begin{eqnarray}\label{lamW}
&&E_h^0={W^2+m_h^2-m_u^2 \over 2 W}, \;
p_h^0={\sqrt{\lambda_W^0} \over 2 W}, \;
\nonumber \\&&
\lambda_W^0=(W^2-m_h^2-m_u^2)^2-4m_h^2W^2.
\end{eqnarray}
All invariants defined via the measured 4--momentum of hadron $p_h$ 
are also different for Born and radiative cases:
\begin{eqnarray}\label{v1v2st}
V_{1,2}^0&=&2k_{1,2}p_h=2(E_{1,2}E_h^0-p_{1,2}p_h^0(\cos\theta_h\cos\theta_{1,2}+
\nonumber \\ && \qquad\qquad
\sin\theta_h\sin\theta_{1,2}\cos\phi_h)),
\nonumber \\
S_t^0&=&2pp_h=W^2+m_h^2-m_u^2+V_2^0-V_1^0.
\end{eqnarray}
The last expression follows from 4--momentum conservation.

The Born cross section of
exclusive pion electroproduction in terms of the introduced variables reads:
\begin{eqnarray}
d\sigma_0 &=& {M_0^2 \over 2S(2\pi)^5}
{d{\vec k_2}\over 2E_2}
{d{\vec p_u}\over 2E_u}
{d{\vec p_h}\over 2E_h} \delta(\Lambda-p_u)=
\nonumber\\
&& ={(4\pi\alpha)^2\over 2(4\pi)^4S^2}\;{L^0_{\mu\nu}W_{\mu\nu} \over
Q^4}\;
{\sqrt{\lambda^0_W} \over W^2} dQ^2dW^2d\Omega_h,
\end{eqnarray}
where $\Lambda$ is the total 4--momentum of the undetected particles,
$\Lambda=p+k_1-k_2-p_h$. For the Born process (Fig.\ref{feyn}a), $\Lambda$ is
equal to the momentum of the undetected hadron, $p_u$, while for the
radiative process (Fig.\ref{feyn}b,c), it is a sum of the undetected hadron
and bremsstrahlung photon momenta, $\Lambda=p_u+k$.  

The phase space is calculated as
\begin{equation}
{d{\vec k_2}\over 2E_2}= \frac{\pi}{2S}dQ^2dW^2
\end{equation}
and
\begin{equation}
{d{\vec p_u}\over 2 E_u}
{d{\vec p_h}\over 2 E_h} \delta(\Lambda-p_u)
=\frac{p_h^0}{4W}d\Omega_h
=\frac{\sqrt{\lambda^0_{W}}}{8W^2}d\Omega_h.
\end{equation}
To calculate the matrix element squared, one needs to 
contract the leptonic and hadronic tensors, 
\begin{equation}
M_0^2={e^4\over Q^4} L^0_{\mu\nu}W_{\mu\nu} = {2e^4 \over
Q^4}\sum_{i=1}^5
\theta^0_i {\cal H}_i^0.
\end{equation}
We consider the longitudinally polarized lepton beam.
In this case the leptonic tensor reads:
\begin{equation}
L_{\mu\nu}^0=\frac{1}{2}{\rm Tr} (\hat k_2+m)\gamma_\mu
(\hat k_1+m)(1+i\gamma_5\hat\xi)\gamma_\nu.
\end{equation}
Here the lepton polarization vector is kept in a general form. If
the lepton is longitudinally polarized and its helicity is positive, 
then the vector $\xi$ can be expressed as
\cite{ASh}
\begin{equation}
\xi={1\over \sqrt{\lambda_s}}\bigl(\frac{S}{m}\; k_1-2m\;p\bigr).
\end{equation}
In the Born approximation, the second term can be dropped. However, 
it gives a non--zero contribution to RC.

For the hadronic tensor, we use a general covariant form
\begin{eqnarray}\label{003}
W^{\mu\nu}&=&-{\tilde g}^{\mu\nu}{\cal H}_1
\;+\;{\tilde p}^{\mu}{\tilde p}^{\nu}{\cal H}_2
\;+\;{\tilde p_h}^{\mu}{\tilde p_h}^{\nu}{\cal H}_3
\\ && \qquad
+ \;( {\tilde p}^{\mu}{\tilde p_h}^{\nu}
  +{\tilde p_h}^{\mu}{\tilde p}^{\nu}
     ){\cal H}_{4}
\;+\;(
 {\tilde p_h}^{\mu}{\tilde p}^{\nu}
-{\tilde p}^{\mu}{\tilde p_h}^{\nu}
     ){\cal H}_{5},
\nonumber
\end{eqnarray}
where the tilde for an arbitrary 4--vector $a^\mu$ denotes the substitution 
${\tilde a}^\mu = a^\mu - \frac{aq}{q^2}q^\mu$ (to ensure electromagnetic gauge invariance). 
The first four structure functions have tensor coefficients symmetric over Lorentz
indices, but the last one is antisymmetric, it 
contributes to the polarization--dependent part of the cross section. The quantities
$\theta_i^0$ have the following form,
\begin{eqnarray}\label{theta0}
  \theta_1^0&=&Q^2,			  \\
  \theta_2^0&=&\frac12 (SX-M^2Q^2),	 \nonumber\\
  \theta_3^0&=&\frac12(V_1^0V_2^0-m_h^2Q^2),  \nonumber\\
  \theta_4^0&=&\frac12(SV_2^0+XV_1^0-S_t^0Q^2), \nonumber\\
  \theta_5^0&=& -2\epsilon(k_1,k_2,p,p_h),\nonumber
\end{eqnarray}
where the operator in the last line is defined as
\begin{equation}\label{gramm}
\epsilon(p_1,p_2,p_3,p_4)=
p_1^\alpha p_2^\beta
p_3^\gamma p_4^\sigma\epsilon_{\alpha\beta\gamma\sigma}.
\end{equation}
In the chosen c.m. frame it is equal to
\begin{eqnarray}
\epsilon(k_1,k_2,p,p_h)&=&WP_qp_hp_1\sin\theta_1\sin\theta_h\sin\phi_h
\nonumber \\
&=&{\sqrt{\lambda_W^0\lambda} \over 4W} \sin\theta_h\sin\phi_h.
\end{eqnarray}

As a result, we obtain for the Born cross section,
($\sigma_0 =d\sigma / dW^2dQ^2d\Omega_h$)
\begin{equation}\label{born}
\sigma_0={ \alpha^2 \sqrt{\lambda_W^0} \over 32\pi^2 S^2W^2Q^4}
\sum_{i=1}^5 \theta^0_i {\cal H}_i^0.
\end{equation}

The expression for the  Born cross section in the form (\ref{born}) is 
convenient for our further calculation. It is equivalent to the
well known formula in terms of photoabsorption cross sections (or response functions),
where each term corresponds to certain polarization states of the virtual photon \cite{dombey},
\begin{eqnarray}\label{response}
\frac{1}{N_0}\sigma_0&=&\sigma_T + \epsilon\;\sigma_L + \epsilon\;
\cos 2\phi_h\;\sigma_{TT}
\nonumber \\ &&\qquad\qquad
+ \sqrt{\epsilon(1+\epsilon)/2}\;\cos\phi_h \; \sigma_{LT}
\nonumber \\ &&\qquad\qquad
+ h_e\sqrt{\epsilon(1-\epsilon)/2}\;\sin\phi_h \; \sigma_{LT}',
\end{eqnarray}
where
\begin{equation}
	N_0=\alpha{W^2-M^2 \over 2S^2Q^2(1-\epsilon)}, 
\end{equation}
$h_e$ is the longitudinal polarization degree of the incoming leptons,
and $\epsilon$ \cite{dombey} describes the virtual photon polarization.

The following expressions relate the structure functions 
(\ref{003}) and Born photoabsorption cross sections (\ref{response}), 
\begin{eqnarray}\label{SFs}
{\cal H}_1(W^2,Q^2,t)&=&C(\sigma_T-\sigma_{TT}),
\nonumber\\
{\cal H}_2(W^2,Q^2,t)&=&{2C\over\lambda_q}
\biggl(2Q^2(\sigma_T-\sigma_{TT}+\sigma_{L}),
\nonumber\\&&
-TQ\sigma_{LT}
+T^2\sigma_{TT}\biggr),
\nonumber\\
{\cal H}_3(W^2,Q^2,t)&=&{2C\lambda_q \over \lambda_l} \sigma_{TT},
\nonumber\\
{\cal H}_4(W^2,Q^2,t)&=&{C\over
\sqrt{\lambda_l}}(2T\sigma_{TT}-Q\sigma_{LT}),
\nonumber\\
{\cal H}_5(W^2,Q^2,t)&=&
{C\over \sqrt{\lambda_l}}Q\sigma_{LT}',
\end{eqnarray}
where
\begin{eqnarray}
C&=&{16\pi^2(W^2-M^2)W^2\over \alpha\sqrt{\lambda_W^0}},
\nonumber\\
T&=&{S_xt_q-2Q^2S_t \over \sqrt{\lambda_l}},
\nonumber\\
\lambda_l&=&Q^2S_t^2-S_tS_xt_q-M^2t_q^2-m_h^2\lambda_q,
\end{eqnarray}
and $Q=\sqrt{Q^2}$.
Note that both structure functions ${\cal H}_i$, as well as cross
sections $\sigma$'s, are functions of
three independent invariant variables, which usually are chosen as $Q^2$,
$W^2$ and $t$. Therefore their transformation coefficients
depend only on these variables but not on $\phi_h$. For the radiative
case the variables will be defined in the next section (see
Eq.(\ref{tildeQWt})). For the Born case,
they are taken as: $W^2$, $Q^2$ and $t=t_0=V_2^0-V_1^0-Q^2+m_h^2$.

\section{Exact formulas for radiative correction}

The cross section of radiative process
is given by
\begin{eqnarray} \label{start}
d\sigma_r &=& {M_r^2 \over 2S(2\pi)^8}
{d{\vec k_2}\over 2E_2}
{d{\vec k}\over 2\omega }
{d{\vec p_u}\over 2E_u}
{d{\vec p_h}\over 2E_h} \delta(\Lambda-k-p_u)
\\
&=& {(4\pi\alpha)^3
dQ^2dW^2d\Omega_h
\over 2(4\pi)^7S^2W^2}
\int d\Omega_kdv {v\sqrt{\lambda_W}\over f_W^2 \tilde
Q^4} L_{\mu\nu}^RW_{\mu\nu},
\nonumber
\end{eqnarray}
where $\Omega_{k(h)}$ stands for the solid angle of the bremsstrahlung
photon (detected hadron).

Let us first  consider the phase space for the radiative process.
Integrating over the 3-momentum of unobserved hadron and using
Dirac $\delta$-function  to eliminate integration over the hadron
energy, we have
\begin{eqnarray}\label{gamphassp}
&&\int{d{\vec k}\over 2\omega }
{d{\vec p_u}\over 2E_u }
{d{\vec p_h}\over 2E_h} \delta(\Lambda-k-p_u)= \qquad
 \nonumber \\ && \qquad
=\frac{1}{64} \int d\Omega_h d\Omega_k \int\limits_0^{v_m} {vdv \over
f_W^2} \sqrt{\lambda_W},
\end{eqnarray}
where
\begin{eqnarray}
f_W&=&W-E_h+p_h(\cos\theta_h\cos\theta_k
\nonumber \\
&&\qquad\qquad
+\sin\theta_h\sin\theta_k
\cos(\phi_h-\phi_k) ).
\end{eqnarray}
Here we introduce a quantity $v$ that describes the missing
mass (or inelasticity) due to emission of a bremsstrahlung photon, 
$v=\Lambda^2-m_u^2$. Note that $v=0$ for the Born process (Fig.\ref{feyn}a),
as well as in the soft--photon limit ($k=0$).
As can be seen from (\ref{gamphassp}), both the
photon and hadron energies are now functions of $v$.
It is related to another quantity, $f_W$, as follows:
\begin{equation}\label{vvv}
v=W^2+m_h^2-m_u^2-2WE_h=2\omega f_W.
\end{equation}
The largest value of inelasticity allowed by kinematics ($v_m$) corresponds to
the threshold of electroproduction. It is therefore defined from
the relation $E_h=m_h$, yielding
\begin{equation}\label{vmax}
v_m=(W-m_h)^2-m_u^2.
\end{equation}
Note that $v_m$ is always smaller for the heavier hardon, namely, the nucleon, detected
in the final state. It does not depend on photon angles, therefore the integration region is
a rectangle.

The maximum inelasticity $v_m$ is an important quantity for the RC calculation. 
All kinematic cuts made by experimentalists in data analysis influence RC. It is often
possible to reduce all these cuts to one effective cut
on the inelasticity (or missing mass), $v_{cut}$.  In which case $v_{cut}$ should replace
$v_m$ as the upper limit of integration in Eq.(\ref{gamphassp}), and thus RC
 can be calculated within the cuts using the obtained formulas. If no cuts are applied, the
maximum value of $v$ equals $v_m$, as given by energy--momentum conservation (\ref{vmax}).

Now we can fix the kinematics of the radiative process. We have to
express all scalar products and kinematic variables in terms of seven
variables: Four variables that define the differential cross
section and three integration variables. First, note that all definitions
of leptonic variables given in (\ref{lepkin}) hold in this
case. Hadron and real photon energies are defined by (\ref{vvv}). Hadron
momentum and scalar products are given as
\begin{eqnarray}\label{radkin}
p_h&=&{\sqrt{\lambda_W} \over 2 W}, \;
\lambda_W=(W^2-m_h^2-m_u^2-v)^2-4m_h^2W^2,
\nonumber \\
V_{1,2}&=&2k_{1,2}p_h=2(E_{1,2}E_h-p_{1,2}p_h(\cos\theta_h\cos\theta_{1,2}+
\nonumber \\ && \qquad\qquad
\sin\theta_h\sin\theta_{1,2}\cos\phi_h)),
\nonumber \\
S_t&=&2pp_h=W^2+m_h^2-m_u^2+V_2-V_1-v
\nonumber \\ && \qquad
=S_x+t+M^2-m_u^2-v,
\nonumber \\
t&=&V_2-V_1-Q^2+m_h^2.
\end{eqnarray}
As in (\ref{v1v2st}), the last expression is obtained from
conservation laws. All hadron kinematic variables depend on inelasticity
$v$ but not on photon angles $\Omega_k$. Scalar products containing
the photon
4--momentum read
\begin{eqnarray}\label{kappa}
2pk&=&R_w(1-\tau)=2\omega(E_p-p_p\cos\theta_k),
 \\
2k_{1,2}k&=&\kappa_{1,2}=2\omega(E_{1,2}-p_{1,2}(\cos\theta_k\cos\theta_{1,2}+
\nonumber \\ && \qquad\qquad
\sin\theta_k\sin\theta_{1,2}\cos\phi_k),
\nonumber \\
2p_hk&=&\mu=2\omega(E_h-p_h(\cos\theta_k\cos\theta_h+
\nonumber \\ && \qquad\qquad
\sin\theta_k\sin\theta_h\cos(\phi_k-\phi_h)).
\nonumber
\end{eqnarray}
Here we use also invariant variables $R_w=2k(p+q)$ and $\tau=2kq/R_w$;
$R_w=f_Wv/W$.

The leptonic tensor of the radiative process has a more complicated form. It can be
written as
\begin{eqnarray}\label{leprad}
L_{\mu\nu}^R&=&\frac{1}{2}{\rm Tr} (\hat k_2+m)\Gamma_{\mu\alpha}
   (\hat k_1+m)(1+i\gamma_5\hat\xi)\hat\Gamma_{\alpha\nu},
\nonumber\\
  \Gamma_{\mu\alpha}&=&
  \biggl[\biggl( {k_{1\alpha}\over kk_1} -
{k_{1\alpha}\over kk_1}\biggr)\gamma_\mu
-{\gamma_\mu\hat k\gamma\alpha \over 2kk_1}
-{\gamma_\alpha\hat k\gamma\mu \over 2kk_2}\biggr],
\nonumber\\
  \hat\Gamma_{\alpha\nu}&=&
  \biggl[\biggl( {k_{1\alpha}\over kk_1} -
{k_{1\alpha}\over kk_1}\biggr)\gamma_\mu
-{\gamma_\alpha\hat k\gamma\nu \over 2kk_1}
-{\gamma_\nu\hat k\gamma\alpha \over 2kk_2}
\biggr].
\end{eqnarray}
After contraction of the radiative leptonic tensor, we obtain
\begin{equation}
M_R^2=-{2e^6\over \tilde Q^4} L_{\mu\nu}^RW_{\mu\nu} =
-{2e^6\over \tilde Q^4 R_w}
\sum_{i=1}^5 \theta_{i}{\cal H}_i.
\end{equation}
$R_w$ was extracted explicitly in order to cancel $v$ coming from
Jacobian (\ref{gamphassp}).
Arguments of ${\cal H}_i$ can be expressed as
\begin{eqnarray}\label{tildeQWt}
\tilde Q^2&=&-(q-k)^2=Q^2+R_w\tau,
\\
\tilde W^2&=&(p+q-k)^2=W^2-R_w,
\nonumber \\
\tilde t&=&(q-k-p_h)^2=t-R_w(\tau-\mu).
\end{eqnarray}
It is well known that the cross section of the radiative process is 
infrared--divergent, which requires careful consideration in order to cancel in the 
difference.
The
procedure will be discussed in the next section. Here we can extract the
infrared convergent terms in $\theta_i$ in separate pieces:
\begin{eqnarray}\label{FIR}
\theta_i&=&\frac{4}{R_w}F_{IR}\theta_i^B+\theta_i^F.
\end{eqnarray}
The quantities $\theta^B_i$ are defined by expressions for $\theta_i^0$
with a reservation that the hadronic quantities $V_1,V_2,S_t$ and vector $p_h$ itself have to be
calculated for the radiative kinematics (\ref{radkin}). This term originates
from the first terms of
$\Gamma_{\mu\alpha}$ and $\bar\Gamma_{\alpha\nu}$
 in definition of (\ref{leprad}).

The explicit form of finite parts of these functions $\theta^F_i$ are given
in Appendix.

Finally, the cross section of the radiative process
($\sigma_R=d\sigma_R/dQ^2dW^2d\Omega_h$) is given as
\begin{equation}\label{sigr}
\sigma_R=-{\alpha^3 \over 2^9\pi^4S^2W^4}\int d\Omega_k dv
{\sqrt{\lambda_W}\over f}\sum_i
{\theta_{i} {\cal H}_i \over \tilde Q^4},
\end{equation}
where $f=f_W/W$.

As a cross--check, we consider the soft--photon limit
$\omega_{min}<R/2M<\omega_{max}\ll$ {\it all energies and masses}.
In this case only the first term in (\ref{FIR}) survives.
\begin{equation}
\sigma_R={2\alpha \over
\pi}\biggl(\log\frac{Q^2}{m^2}-1\biggr)\log\frac{w_{max}}{w_{min}}
\sigma_0.
\end{equation}
Integration over photonic angles is performed analytically,
\begin{equation}
 \int d\Omega_k F_{IR} = -2\bigl( l_m
-1\bigr),\quad l_m=\log \frac{Q^2}{m^2}.
\label{ap10} \end{equation}

\section{Infrared divergence}

As was already mentioned, this cross section contains an infrared divergence. 
Therefore, in order to compute RC, we have to use some regularization method first.
We use the method of Bardin and Shumeiko for
a covariant treatment of an infrared divergence problem. Basically, we follow the
original papers devoted to this topic, namely, Ref. \cite{BSh} for $e\mu$ elastic
scattering and Ref. \cite{Ak} where exclusive
electroproduction was considered. One can find a good and detailed review in
Ref. \cite{ABK}.

Following the rules of dimensional regularization,
we apply an identity transformation to the radiative cross
section assuming that we deal with $n$-dimensional space:
\begin{equation}
\sigma_R = \sigma_R - \sigma_{IR}+\sigma_{IR}
=\sigma_F+\sigma_{IR}.
 \end{equation}
The purpose of the procedure is to separate the cross section into two pieces.
The first term $d\sigma_F$ is complicated but infrared free. Infrared
divergence is contained in the separate term $d\sigma_{IR}$, which has a quite simple
structure and can be analytically calculated within $n$-dimensional space.
In principle, there is some arbitrariness in choosing the form of the
subtracted term. Actually only the asymptotical form for $R_w$ (or
$w$) $\rightarrow 0$ is fixed. Another limitation comes from the theorem
about a possibility to switch the order of integration and the limit
$n\rightarrow 4$. It means that we have to provide uniform
convergence of $d\sigma_F$ in the limit. Practically, it is ensured
if the subtracted term has a structure $F/w$, and the difference is 
$(F(w)-F(0))/w$.

We define the subtracted part of the radiative cross section as follows:
Only the first term from Eq. (\ref{FIR}) r.h.s. is kept. It gives the required $1/R_w$ behavior of
the radiative cross section. Everywhere else, except for the $\delta$-function, we assume
$v=0$. This allows to factorize the Born cross section when
calculating the correction:

\begin{equation}\label{sigir}
\sigma_{IR}=-\sigma_0{\alpha \over \pi^2}\int dv
\int{d\vec k \over \omega} F_{IR}
\delta((\Lambda-k)^2-m_u^2).
\end{equation}

In the cross section $\sigma_F$ we can now remove the regularization.
The result is
\begin{eqnarray}
\sigma_F&=&-{\alpha^3 \over 2^9\pi^4S^2W^2}\int d\Omega_k {dv \over f}
\times
\\ && \qquad
\sum_i
\biggl[
{\sqrt{\lambda_W} \over \tilde Q^4}\theta_{i} {\cal H}_i
-{4F_{IR} \sqrt{\lambda_W^0}\over  Q^4}\theta_{i}^0 {\cal H}_i^0\biggr].
\nonumber
\end{eqnarray}

For calculation purposes, we split the integration region into two parts
separated by the infinitesimal value of inelasticity $\bar v$:
\begin{equation}
\sigma_{IR} = \frac{\alpha}{\pi}\delta_R^{IR}\sigma_0
= \frac{\alpha}{\pi}(\delta_S+\delta_H)\sigma_0
 \end{equation}
with
\begin{eqnarray}
\delta_{S}&=&\frac{-1}{\pi}\int\limits_0^{\bar
v}dv\int\frac{d^{n-1}k}{(2\pi\mu)^{n-4}k_0}F_{IR}\delta((\Lambda-k)^2-m_u^2),
\nonumber \\
\delta_{H}&=&\frac{-1}{\pi}\int\limits_{\bar
v}^{v_m}dv\int\frac{d^3k}{k_0}F_{IR}\delta((\Lambda-k)^2-m_u^2).
\end{eqnarray}
The term $\delta_S^{IR}$ corresponds to the soft photon contribution, while
the term $\delta_H^{IR}$ is caused only by hard photons and therefore
it does not contain the infrared singularity. 
Again, in the second contribution the regularization can be removed.

We keep integration over the 3--momentum of the radiated photon in a covariant form.
It makes it possible to calculate the integral in any frame.
We choose the frame where $\vec\Lambda =0$ (so--called R--frame).
Integration over inelasticity is external, therefore $v$ is fixed for the
integral. 

Calculation of the hard--photon contribution is straightforward,
\begin{equation}
\delta_H=(l_m-1)\log\frac{v_m}{\bar v}.
\end{equation}

For $\delta_S$, we follow Ref. \cite{BSh} generalizing these
calculations to the exclusive electroproduction case.

Using the spherical $n$-dimensional frame, we have \cite{BSh,Sirlin,March}
\begin{eqnarray}
&&\delta_v=\frac{1}{\pi}\int\limits_0^{v_m}dv
{2\pi^{n/2-1}\over (2\pi\mu_0)^{n-4}\Gamma(\frac{n}{2}-1)}
\int dk_0 k_0^{n-3} \times
\nonumber \\ &&
\times\int
 \sin^{n-3}\theta
 d\theta
\biggl( \frac{k_1}{2k_1k} -\frac{k_2}{2k_2k} \biggr)^2
 \delta(v-2k_0m_u).
\end{eqnarray}

\begin{eqnarray}\label{FAXI}
&&F(\alpha,cos(\theta))=\biggl( \frac{k_1}{2k_1k} -\frac{k_2}{2k_2k}
\biggr)^2
=
\frac{1}{4k_0^2}\biggl[
 {m^2 \over E_{1R}^2(1-\beta_1\cos\theta)^2}
\nonumber \\&&
+{m^2 \over E_{2R}^2(1-\beta_2\cos\theta)^2}
-\int\limits_0^1{d\alpha Q^2 \over
E_\alpha^2(1-\beta_\alpha\cos\theta)^2
} \biggr]
\end{eqnarray}
Here Feynman  parameters are introduced in order to join two denominators. Thus we define
the new vector
\begin{equation}
k_\alpha=\alpha k_1+(1-\alpha)k_2
\end{equation}
and quantities $\beta_{1,2,\alpha}$ as a ratio of energy to momentum in 
the R--frame. In terms of invariants, we have
\begin{eqnarray}
&&\beta_1=\sqrt{1+{4m^2m_u^2\over {S'_0}^2}},\quad
\beta_2=\sqrt{1+{4m^2m_u^2\over {X'_0}^2}},\quad
\nonumber \\ &&
\beta_\alpha=\sqrt{1+{4m_u^2(m^2+\alpha(1-\alpha)Q^2)\over (\alpha
S'_0+(1-\alpha)X'_0)^2}}.
\end{eqnarray}
It should be noted  that the same polar angles are used in Eq. (\ref{FAXI}) for
all three terms. That is why we can rotate the coordinate system for all the
terms and choose the $OZ$ axis along $\vec k_{1,2,\alpha}$, respectively. After straightforward
integration over the polar angle, we obtain Eq. (\ref{FAXI}).

The next two steps involve integration over $k_0$ using 
the $\delta$-function and over $v$
\begin{equation}
\int\limits_0^{\bar v} dv \int dk_0
k_0^{n-5}\delta(v-2k_0m_u)=\frac{1}{n-4}\biggl(\frac{\bar
v}{2m_u}\biggr)^{n-4}.
\end{equation}
Since the pole $(n-4)^{-1}$ is extracted, we can use expansion in series of
$n-4$,
\begin{eqnarray}
&&
\frac{1}{n-4}
\biggl({\bar v \over 4\pi\mu_0m_u}\biggr)^{n-4}
{(\pi(1-xi^2))^{n/2-2} \over \Gamma(\frac{n}{2}-1)}
 \\ && \qquad \qquad
\rightarrow
P_{IR}+\log{\bar v \over 2\mu_0m_u}
+\log\sqrt{1-\xi^2}.
\nonumber
\end{eqnarray}
with $\xi=cos\theta$ and standard $P_{IR}$ defined as
\begin{equation}
P_{IR}=\frac{1}{n-4}+\frac{1}{2} \gamma_E+\ln \frac{1}{2 \sqrt{\pi}},
\end{equation}
$\gamma_E$ being the Euler constant.
After then we reduce the expression for $\delta_S$ to the form
\begin{equation} 
\delta_S=\int d\alpha d\xi\biggl[
P_{IR}+\log{\bar v \over 2\mu_0m_u}+\log\sqrt{1-\xi^2}
\biggr]F(\alpha,\xi),
\end{equation}
which includes only standard integration that can be done using, for instance, tables from Appendix
D of Ref.\cite{ABK}. Integration is straightforward, but it contains
an integral usually associated with a so--called $S_{\phi}$ function in
the Bardin and Shumeiko approach. Here we skip the discussion about properties
of the function and give the result in the ultra--relativistic approximation
which we use for the entire calculation:
\begin{eqnarray}
S_\phi&=&{Q^2 \over 2} \int\limits_0^1 {d\alpha \over
\beta_{\alpha}(m^2+\alpha(1-\alpha)Q^2)}
\log{1-\beta_\alpha \over 1+\beta_\alpha}=
\nonumber \\ && \qquad	\qquad
\frac{1}{2}l_m^2-l_m\log{S'_0X'_0\over m^2m_u^2}
-\frac{1}{2}\log^2{S'_0\over X'_0}
\nonumber \\ && \qquad \qquad \qquad
+{\rm Li}_2\biggl(1-{Q^2m_u^2\over S'_0X'_0}\biggr)
-\frac{\pi^2}{3}.
 \end{eqnarray}

Finally, we have for $\delta_S$
\begin{equation}
\delta_S=2\biggl(P_{IR}+\log\frac{\bar v}{\mu
M}\biggr)(l_m-1)+\log\frac{S'X'}{m^2M^2}+S_{\phi}.
\end{equation}

The infrared--divergent terms $P_{IR}$, as well as the parameters $\mu$ and $\bar v$ are completely
canceled in the sum  $\delta_S +
\delta_H$ with $\delta_V$ which is a contribution of the vertex function (Fig.\ref{feyn}d):
\begin{equation}
\delta_V=-2(P_{IR}+\log\frac{m}{\mu})(l_m-1)-\frac{1}{2}l_m^2+\frac{3}{2}l_m
-2+\frac{\pi^2}{6}.
\end{equation}
For this sum, we have
\begin{equation}
\frac{\alpha}{\pi} (\delta_S+\delta_H+\delta_V)=\delta_{inf}+\delta_{VR},
\end{equation}
where
\begin{eqnarray}	\label{deltas}
\delta_{VR} &=&\frac{\alpha}{\pi}
\biggl(\frac{3}{2}l_m-2-\frac{1}{2}\ln^2\frac {X'_0}{S'_0}+{\rm Li}_2
\biggl[1-\frac{Q^2M^2}{S'_0 X'_0}\biggr]-\frac{\pi^2}6\biggr),
\nonumber\\[0.2cm]
\delta _{inf}&=& \frac{\alpha}{\pi}
(l_m-1)\ln\frac{v_m^2}{S'_0X'_0}.
\end{eqnarray}
Higher--order corrections can be partially taken into account using a
special procedure of exponentiation of multiple soft photon radiation.
There is an uncertainty about which part of $\delta_{VR}$ has to be
exponentiated. Within the approach \cite{Sh}, we have to change
$1+\delta_{inf}$ to $\exp \delta_{inf}$.

Collecting all the terms,  we obtain
for the case of meson electroproduction:
\begin {equation}
\sigma _{obs} = \sigma _0 e^{\delta_{inf}}
(1+ \delta_{VR}+\delta_{vac})+\sigma_{F}.
\label{eq1}
\end {equation}
Here the corrections
 $\delta_{inf}$  and   $\delta_{vac}$  come from  radiation of soft
photons and effects of vacuum polarization,
the correction	$\delta_{VR}$ is an infrared--free sum of factorized
parts of real and virtual photon radiation, and $\sigma_F$ is an 
infrared--free contribution from the bremsstrahlung process.

\section{Leading Log Approximation}

In this section we extract the leading log contribution from formulas
obtained in the previous section. After then we show that the result
coincides with what we obtain from a generalized leading log
approximation.

The method of extraction of leading log (or peaking) contributions
was first suggested in Ref. \cite{MoTsai}
(see also papers \cite{Sh,approx}).
Formally, we have to calculate residues of the terms $1/z_1$ and
$1/z_2$. The corresponding poles appear in $\tau$  for
$\tau=\tau_s=-Q^2/S$ and $\tau=\tau_x=Q^2/X$. As a result, we have (apart
from the factorizable correction) two contributions of hard radiation,
\begin{equation}\label{llcrsec}
\sigma_{LL}={\alpha \over 2\pi} (l_m-1)
\biggl(
\Bigl(
      3    +2\log{v_m^2\over S'X'}\Bigr)\sigma_0+\sigma_S+\sigma_X
\biggr).
\end{equation}

Technically, the contributions can
be obtained as
\begin{eqnarray}\label{ll1}
 \int d\Omega_k \;\theta_i
&=&-8\pi l_m{W^2 \over S-Q^2}
\;{1+z_1^2 \over z_1(1-z_1)}\; \theta_{i}^B,
\nonumber\\
 \int d\Omega_k \;\theta_i
&=&-8 \pi l_m
{W^2 \over X+Q^2} \;
{1+z_2^2 \over 1-z_2} \; \theta_{i}^B.
\end{eqnarray}
For the two peaks the contributions are, respectively:
\begin{eqnarray}\label{ll2}
\sigma_s&=&\int\limits_{0}^{v_m} {dv \over S' }
\biggl({1+z^2_1 \over 1-z_1 } C_s \sigma_s
-{2\sigma_0 \over 1-z_1}\biggr),
\nonumber \\
\sigma_x&=&\int\limits_{0}^{v_m} {z_2dv \over X' }
\biggl({1+z^2_2 \over 1-z_2 } C_x \sigma_x
-{2\sigma_0 \over 1-z_2}\biggr),
\end{eqnarray}
where $z_1=1-v/S'$, $z_2=X'/(X'+v)$ and
\begin{eqnarray}
&&C_{s,x}=
{\sqrt{\lambda_W} \over
\sqrt{\lambda^0_{W\;s,x}}}{W_{s,x}^2 \over W^2},
\nonumber \\
&& W^2_s=z_1S-X-z_1Q^2-M^2,
\nonumber\\&&
 W^2_s=S-{X \over z_2}-{Q^2 \over z_2}-M^2,
\end{eqnarray}
and $\lambda^0_{W\;s,x}$ are calculated in accordance with Eq. (\ref{lamW})
using $W^2=W^2_{s,x}$.

These leading log formulas were extracted from our exact 
expressions given in the previous section. The result
can also be obtained using standard leading log techniques \cite{DeRujula} from
our expression (\ref{start}). The leading term comes from
angular integration of denominators like $kk_1$ and $kk_2$. They can be
extracted at the level of leptonic tensor
\begin{equation}\label{sta1}
L_{\mu\nu}^R=L_{\mu\nu}\biggl[
{1+z_1^2\over 1-z_1 }{1\over kk_1}+
{1+z_2^2\over z_2(1-z_2) }{1\over kk_2} \biggr].
\end{equation}
Only these scalar products in the denominators are subject to
angular integration. For our exclusive process, integration yields:
\begin{equation}\label{sta2}
\int {d \vec p_u \over 2\epsilon_u}  {d \vec k \over 2\omega}
\delta(\Lambda-k-p_u) {1\over kk_{1,2}}=\biggl[
{\pi l_m \over S'},
{\pi l_m \over X'}
\biggr].
\end{equation}
Now, using
(\ref{start}), (\ref{sta1}) and (\ref{sta2}), we obtain the following
leading--log cross section for the radiative process:
\begin{eqnarray}\label{llog}
{d^6\sigma \over d \Gamma^6 }&=&-{\alpha l_m \over 2\pi}
\biggl[
\frac{1}{S'}{1+z_1^2\over 1-z_1 } {d^6\sigma_0(z_1k_1,k_2) \over d
\tilde \Gamma^6}+
\nonumber\\ && \qquad \quad
+\frac{1}{X'}{1+z_2^2\over 1-z_2 } {d^6\sigma_0(k_1,k_2/z_2) \over d
\tilde \Gamma^6} \biggr],
\nonumber\\
d \Gamma^6 &=& \frac{d^3 \vec k_2}{2\epsilon_2} \frac{d^3 \vec
p_h}{2\epsilon_h}.
\end{eqnarray}

While deriving the above results, (\ref{ll1},\ref{ll2}), we take
into account relations between $z_{1,2}$, $v$ and $S'$ ($X'$), which
follow from the constraint $(\Lambda^2-(1-z_1)k_1)^2-m_u^2=0$ or
$(\Lambda^2-(1-1/z_2)k_2)^2-m_u^2=0$. They are
\begin{equation}
v=(1-z_1)S', \qquad v={1-z_2 \over z_2} X'.
\end{equation}


\section{Exclusive and inclusive radiative correction}

An important consistency test is to show that inclusive RC can be 
obtained by integration over
hadronic angles. It should be noted that this is not trivial because
the hadronic angles of radiative and Born cross sections are defined even in
different frames. For definition of the Born angles, we use the center-of-mass
frame, while for angles of the radiative process we have to use another frame
defined by vectors $p$ and $q-k$. As a result, it leads to quite complicated
kinematic relations between these angles. 

We start with Eq.(\ref{llcrsec}). After integration over hadronic
angles, we have to obtain the leading log cross section
($\sigma^{inc}_{LL}=d\sigma/dW^2dQ^2$) for inclusive case.
The inclusive formula can be found in Refs. \cite{approx,POLRAD20}, for example. For double
differential cross sections in $Q^2$ and $W^2$, it reads
\begin{equation}\label{llcrsecinc}
\sigma_{LL}^{inc}={\alpha \over 2\pi} (l_m-1)
\biggl(
\Bigl(
      3    +2\log{v_m^2\over
u_1u_2}\Bigr)\sigma_0^{inc}+\sigma_S^{inc}+\sigma_X^{inc},
\biggr)
\end{equation}
with
\begin{eqnarray}\label{sisxinc}
\sigma_S^{inc}&=&{\alpha\over 2\pi}l_m
\int\limits_{z_1^m}^1dz_1\; z_1{1+z_1^2 \over 1-z_1} \sigma_{0S}^{inc}
-{ 2\sigma_{0}^{inc}\over 1-z_1},
\\
\sigma_X^{inc}&=&{\alpha\over 2\pi}l_m
\int\limits_{z_2^m}^1{dz_2\over z_2} {1+z_2^2 \over 1-z_2}
\sigma_{0X}^{inc}
-{ 2\sigma_{0}^{inc}\over 1-z_2},
\end{eqnarray}
where $z_1^m=X/u_2$
and $z_2^m=u_1/S$.

Let us consider integration over $\Omega_h$ of radiative
cross sections ($i.e.$, the first terms on the left--hand--sides in (\ref{sisxinc})). The
simplest way to relate angles in different frames is to express them in
terms of kinematic invariants. We use the following relation,
\begin{equation}
d\Omega_h=\frac{W^2}{\sqrt{\lambda_W^0}} {dV_1^0 dV_2^0 \over
\sqrt{-D(V_1^0,V_2^0)}},
\end{equation}
where $D(V_1^0,V_2^0)=[\epsilon (k_1,k_2,p,p_h)]^2$ is Gramm determinant (see
(\ref{gramm}) and
Ref.\cite{Bukling_Kajanti}).
For the cross sections $\sigma_S^{inc}$ and $\sigma_X^{inc}$ the corresponding
expressions are
\begin{equation}
d\Omega_h=\frac{W^2_s}{\sqrt{\lambda_{Ws}^0}} {z_1dV_1 dV_2 \over
\sqrt{-D(z_1V_1,V_2)}}
\end{equation}
and
\begin{equation}
d\Omega_h=\frac{W^2_x}{\sqrt{\lambda_{Wx}^0}} {dV_1 dV_2 \over
z_2\sqrt{-D(V_1,V_2/z_2)}}.
\end{equation}

The expression for radiative kinematics is
\begin{eqnarray}
16D(V_1,V_2)&=&(u_1V_2+u_2V_1-(W^2+m_h^2-m_u^2-v)Q^2)^2
\nonumber\\ && \;
-4(SX-M^2Q^2)(V_1V_2-m_h^2Q^2).
\end{eqnarray}
The Born case is reached in the limit
$v \rightarrow 0$. Two comments are in order before
explicit integration. The integration area in variables $V_1$ and $V_2$ is
defined by equation $D(V_1,V_2)=0$ that produces an ellipse.
For radiative cross sections $\sigma_S^{inc}$ and
$\sigma_X^{inc}$ this area is $z_1$ (or $z_2$)--dependent. 

A simple  integration procedure follows:
\begin{eqnarray}
&&
\int d\Omega_h \int\limits_0^{v_m} dv
\rightarrow
\int\limits_0^{v_m} dv
\int d\Omega_h
\rightarrow
\int\limits_0^{v_m} dv
\frac{W^2}{\sqrt{\lambda_W}}\int {dV_1dV_2 \over \sqrt{-D}
}
\nonumber \\ &&
\rightarrow
\int dz_1dV_1^sdV_2 \frac{W^2}{\sqrt{\lambda_W}} {S' \over z_1 \sqrt{-D}}
\nonumber \\ &&
\rightarrow
\int dz_1d\Omega_h^s \frac{W^2}{\sqrt{\lambda_W}}
{S' \over z_1 \sqrt{-D}}{ \sqrt{\lambda_{Ws}}\sqrt{-D_s} \over W^2_s }.
\end{eqnarray}
The final formulas for the transformation
\begin{equation}\label{OmOm}
\int d\Omega_h \int\limits_0^{v_m} dv=
\int\limits_{z_1^m}^1 dz_1 \int d\Omega_h^s \frac{S'}{C_s}
\end{equation}
are obtained if we use the equality that
can be checked directly:
\begin{equation}
\frac{D_s}{D}=z_1^2, \qquad
\frac{D_x}{D}=\frac{1}{z_2^2}.
\end{equation}
Using Eq. (\ref{OmOm}), one can see that the exclusive radiative cross section
transforms to the inclusive one after integration.

Let us show how to obtain a factorized part of the inclusive correction
from the exclusive result. The subtracted part of Eqs.(\ref{ll2}) can be
rewritten as
\begin{eqnarray}
\int\limits_0^{v_m} {dv \over S'(1-z_1)}&=&
\int\limits_{z_{1^m}^1}{dz_1\over 1-z_1}
-\int\limits_{S'(v=v_m)}^{S'(v=0)} {dS'\over S'}
 \\ = &&\int\limits_{z_{1^m}^1}{dz_1\over 1-z_1}
-\log\biggl(  {S'_0W\over u_1 (W-m_h)}     \biggr).\nonumber
\end{eqnarray}

After similar calculation for $\sigma_X$,  we can see that inclusive RC is
reproduced exactly.

\section{Numerical analysis}
\label{numerical}

\begin{figure}[t]
\includegraphics[height=8cm,width=8cm]{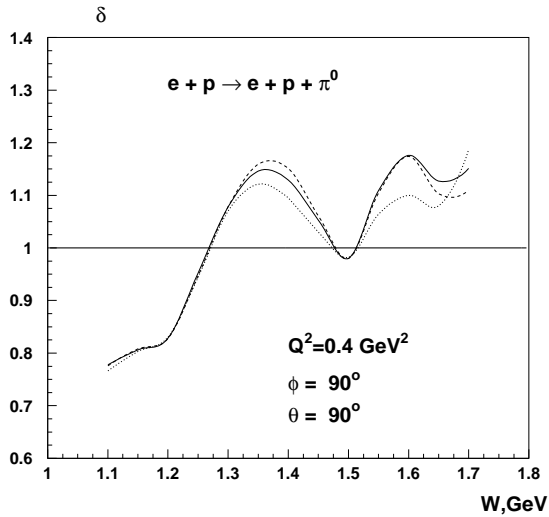}
\caption{\label{wexc1}
$W$-dependence of RC to the cross section of neutral pion production. 
The models used are MAID2000 (solid curve), MAID'98 \cite{drec1} (dashed curve) and AO 
\cite{burkert} (dotted curve).}
\end{figure}

Based on the derived analytical formulas, a Fortran code EXCLURAD was developed. This
code computes RC to the four--fold cross section 
($d^4\sigma /
dWdQ^2d\cos\theta d\phi_h$)
and to polarization beam asymmetry for processes (\ref{process}).  Both exact and leading log formulas obtained
in the previous sections are included. Any value of the inelasticity cut can be optionally chosen.

\begin{figure}[t]
\includegraphics[height=8cm,width=8cm]{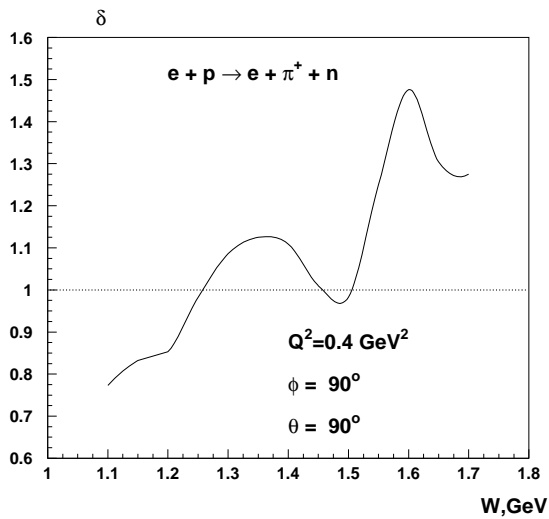}
\caption{\label{wexc2}
$W$-dependence of RC to cross section of charged pion production. 
Kinematics and notation are as in Fig.\ref{wexc1}.}
\end{figure}

Once the computational algorithm for RC is established, there are two possible
ways of incorporating RC into analysis of experimental data on electroproduction:

a) Iteration procedure similar to the one implemented in the RC code POLRAD for inclusive
reactions \cite{POLRAD20} or 

b) Using realistic models for the structure functions (\ref{SFs}) of coincidence electroproduction.
 
Although the former choice seems attractive due to its model independence, it requires, however, 
full and precise experimental mapping of the structure functions (\ref{SFs}) 
in the entire range of the kinematic variables needed to compute the radiative 
process (\ref{sigr}). Such a procedure is much more challenging and less efficient than for 
inclusive processes. Given that such mapping is not available at this time, the only choice
left is (b). In this way, RC is applied to the model calculations, and then model parameters
are fixed from available experimental data. Thus, RC appears as a necessary intermediate step
in extracting model parameters from measurements.   

We use the following models for calculation of structure functions (\ref{SFs}):

{\bf MAID} \cite{drec1}.  
In this model, baryon resonances are described using Breit--Wigner forms,
while background contributions are described using standard Born terms,  
mixed pseudovector--pseudoscalar $\pi$NN coupling and t--channel vector
meson exchange. The final amplitudes are constrained by unitarity and
gauge invariance. We use two versions of MAID: the earlier one, denoted MAID'98
and the most recent, quoted here as MAID2000.

{\bf AO (Amplitude and Observables)}\cite{burkert}.
The amplitudes are parameterized as follows: 
S--channel resonances are parameterized with relativistic 
Breit--Wigner forms with momentum--dependent widths. This part of the
amplitude is complex. In addition, the s--channel and u--channel pion
Born terms are included. These are real numbers. Additional real
background amplitudes are used with an energy dependence that has 
correct threshold behavior.

\begin{figure}[t]
\includegraphics[height=8cm,width=8cm]{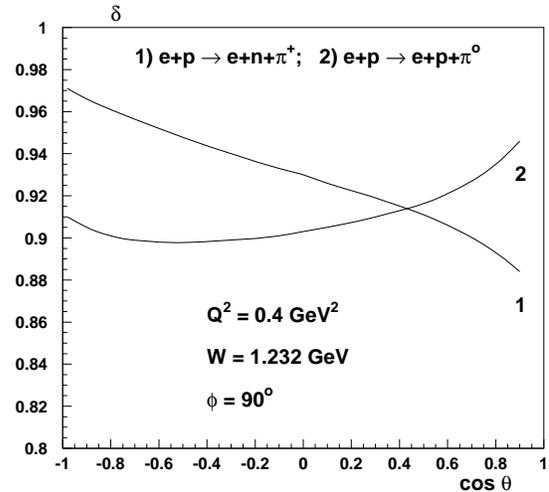}
\caption{\label{aexc}
RC to the cross section  as a function of $\cos\theta$.}
\end{figure}

\begin{figure}[t]
\includegraphics[height=8cm,width=8cm]{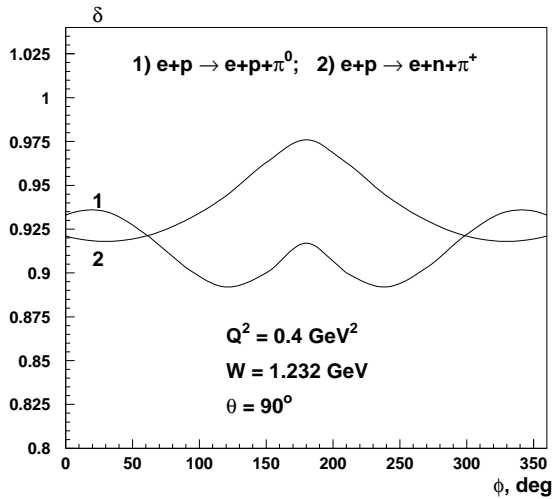}
\caption{\label{pexc}
Dependence of RC to the cross section on the azimuthal angle $\phi$.}
\end{figure}

Another model that can be included into the code EXCLURAD
in a simple and straightforward way is the dynamical model of Sato and Lee \cite{sat01}. 
In this model, the off--shell non--resonant contributions 
to $\gamma^*p\rightarrow\Delta^+(1232)$ were calculated
directly by applying reaction theory within the Hamiltonian
formulation underlying 'bare' photo--coupling form factors. 
It should be noted that the Sato--Lee model does not
include the contributions from higher resonances.

Since the computational algorithm for RC does not depend on a particular choice of
hadronic structure functions, addition of any other model does not constitute a problem.

\begin{figure}[t]
\includegraphics[height=8cm,width=8cm]{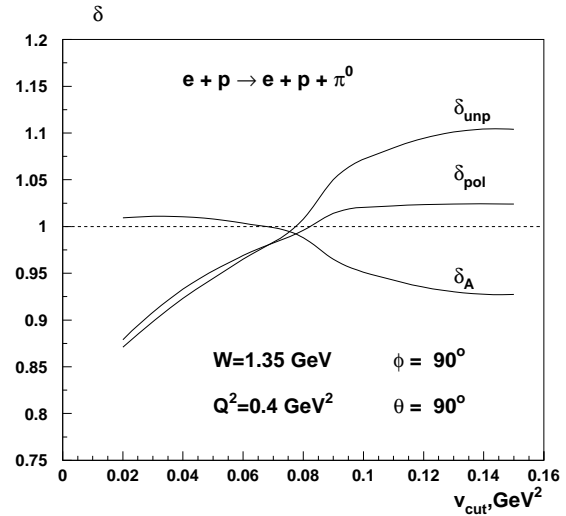}
\caption{\label{cutv}
Dependence of RC to cross section and polarization beam asymmetry on the inelasticity
cut $v$ for neutral pion production. The quantity $\delta_{unp}$ is RC to the unpolarized part of the cross section,
$\delta_{pol}$ is RC to the polarized part of the cross section, and 
$\delta_A=\delta_{unp}/\delta_{pol}$, i.e., it is RC to the beam polarization asymmetry.
MAID2000 was used for structure functions.}
\end{figure}

With the models at hand, we can now evaluate numerically the magnitude of RC
as a function of various kinematics variables. Let us define the RC factor as follows,
\begin{equation}
\delta={\sigma_{obs} \over \sigma_0}.
\end{equation}
For all the following plots, the electron beam energy is $E_{beam}$= 1.645 GeV and
no cuts on inelasticity were used, except for Fig.\ref{cutv}. We choose representative
kinematics for current experiments with the CLAS detector at Jefferson Lab Hall B.
The RC factors to the pion electroproduction cross sections are presented in Figs.\ref{wexc1} 
and \ref{wexc2} as a function of $W$.  In this region, the characteristic features of
the cross section vs. $W$ are strong $\Delta$(1232) and S$_{11}$(1535) resonance peaks. 
One can see that RC is negative at the resonance peaks and positive in the dip regions between
the peaks. Thus it tends to weaken the stength of the peaks and fill the dip regions
above the given resonances. Also shown in Fig. \ref{wexc1}
is the model dependence effect which appears to be noticeable at higher $W$ away from
the resonance peaks. At higher values of $W$, the magnitude of RC is larger in the 
charged pion production case. This is mainly due to the wider range of inelasticity $v$ (\ref{vvv})
associated with detection of a lighter hadron ($i.e.$, the pion). 
    
Before discussing the angular dependence of RC, let us first comment on the
definitions of hadronic angles in order to  avoid possible ambiguities. 
The c.m.s. angles $\theta$ and $\phi$ are between the direction of momentum lost by electrons 
($\vec q=\vec k_1-\vec k_2$) and momentum of the final pion $\vec p_h$, provided
that the pion is detected. For the neutral pion production case, the convention
is to also use the direction of pion momentum reconstructed from kinematics.
In order to follow the convention for the neutral pion production, we define
$\theta$ and $\phi$ with respect to the direction of $-\vec p_h$ {\it opposite} to the 
final proton momentum, while keeping in mind that it is different from the final pion momentum
for the radiative processes Fig.\ref{feyn}b,c.  

The angular dependence of RC is shown in
Figs.\ref{aexc} and \ref{pexc}, where it is plotted as a function of $\cos\theta$ and $\phi$,
respectively.
The kinematics corresponds to the $\Delta$(1232)-peak, where RC leads to
suppression of the cross section. It can be seen from Fig.\ref{aexc} that, approaching
the forward direction,  at the given value of $\phi$, the magnitude
of RC as a function of $\cos\theta$ for the charged--pion channel smoothly increases from 
about 3\% to 12\%, while
for the neutral pions is decreases from 10\% to 5\%. A common feature for both the
channels is that RC is larger in magnitude for the parallel kinematics, when the detected
hadron moves along the transferred momentum. The RC factor varies as a function of azimuthal
angle $\phi$, as well (Fig.\ref{pexc}). Dependence of RC on the angle $\phi$ has important 
implications for the super--Rosenbluth separation of electroproduction
structure functions, and the $\theta$ dependence would affect the partial--wave analysis,
resulting in corrections to electroexcitation parameters of baryon resonances.

\begin{figure}[t]
\includegraphics[height=8cm,width=8cm]{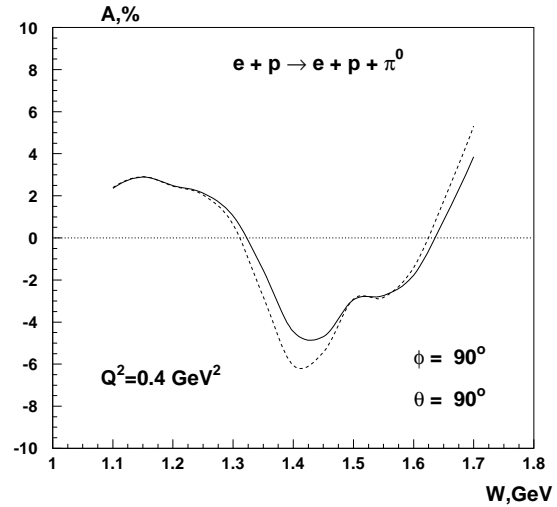}
\caption{\label{excasy1}
$W$-dependence of the beam polarization asymmetry in neutral pion production.
The solid (dashed) curve denote the asymmetry with (without) RC. MAID2000 was used
to compute the structure functions.}
\end{figure}

\begin{figure}[t]
\includegraphics[height=8cm,width=8cm]{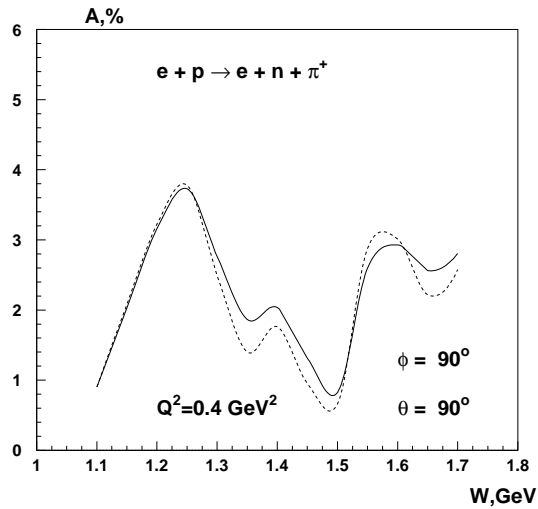}
\caption{\label{excasy2}
The beam polarization asymmetry in charged pion production as a function of $W$ with
(solid curve) and without (dashed curve) RC.
Notation is as in Fig.\ref{excasy1}.}
\end{figure}

The beam polarization asymmetry is plotted in Figs. 
\ref{excasy1} and \ref{excasy2} for the neutral and charged channels, respectively.
One can see that for the asymmetry, RC changes from enhancement to suppression
when passing across the $\Delta$ and S$_{11}$ resonance regions. The RC factor
is most substantial in the dip regions between resonances. 

Figure \ref{cutv} demonstrates the dependence on inelasticity cut $v_{cut}$ (\ref{vvv}).
It can be seen that for the smaller values of the cut, resulting in selection
of softer bremsstrahlung photons, RC to both polarization--dependent
and polarization--independent parts of the cross section is almost the same, resulting
in a small correction to polarization asymmetry. As $v_{cut}$ increases, the correction to the
asymmetry also increases due to the hard--photon emission coming into play.

Any experimental spectrum for electroproduction contains a radiative tail that, due to the finite
energy resolution, cannot be experimentally separated from Born contribution. In practice,
detected events are included into data analysis up to a certain  (cut) limit and 
are interpreted as radiative events.  The contribution of radiative events should be calculated
theoretically using the same inelasticity cut ($v_{cut}$). 
The resulting contribution of the radiative events is subtracted from the 
integrated experimental spectrum in order to obtain the value of Born cross section that,
naturally, should be cut--independent.
  
The Fortran code EXCLURAD can be downloaded from http://www.jlab.org/RC or obtained directly
from the authors.

\section{Discussion and conclusion}
In this paper, we obtain
explicit formulas for the lowest--order QED radiative correction to
cross section and polarization beam asymmetry in the exclusive pion
electroproduction. Analytic formulas are tested in several ways. 
Apart from traditional cross--checks like soft--photon and leading--log limits, it is
found that integration with respect to the hadronic angles reproduces the inclusive 
radiative correction.   

A Fortran code EXCLURAD is developed on the basis of the
analytic formulas. Numerical analysis carried out for JLab kinematic conditions
shows that:
\begin{itemize}
\item  
Radiative correction to the cross section of 
electroproduction is very sensitive to the cut on inelasticity. The harder
cut eliminates contributions from the higher--energy part of the bremsstrahlung spectrum,
thus leading to the smaller magnitude of RC to the polarization asymmetry.

\item RC to cross sections can be as high as several tens of per cent.

\item RC may vary depending on the chosen model for electroproduction
structure functions. The proposed RC procedure may be viewed as a necessary
intermediate step in interpretation of experimental results in terms of
model parameters. Iteration procedure may be required for the regions
where model dependence is substantial.

\item RC have a  non--trivial angular dependence in $\cos \theta$
and $\phi$. This is particularly significant, as these are often used as input
to extract structure functions and partial wave amplitudes.

\end{itemize}

\section*{Acknowledgements}
We would like to acknowledge useful discussions with our 
colleagues at Jefferson Lab. 
We thank Georgii Smirnov for reading the manuscript and useful comments.
This work was supported by the US Department of Energy
under contract DE-AC05-84ER40150. 

\appendix
\section{}




In this Appendix, we give formulas for $\theta_i^F$ (
Eq.(\ref{FIR})). In the unpolarized case,
\begin{equation}
 \theta_i^F = \theta_{i2}+R_w \theta_{i3},
\end{equation}

\begin{eqnarray}
   \theta_{12}&=&4 F_{IR} \tau,
\nonumber\\[0.2cm]
   \theta_{13}&=&-4 F-2 F_{d}, \tau^2,
\nonumber\\[0.2cm]
  2\theta_{22}&=&-4 M^2 F_{IR} \tau-F_{d} S_p^2 \tau+F_{1+} S_p S_x
\nonumber\\ &&
  +2 F_{2-} S_p+2 F_{IR} S_x,
\nonumber\\[0.2cm]
  2\theta_{23}&=&4 M^2 F+2 M^2 F_{d} \tau^2-F_{d} S_x \tau-F_{1+} S_p,
\nonumber\\[0.2cm]
  2\theta_{32}&=&-4 m_h^2 F_{IR} \tau-F_{d} \tau V_+^2+F_{1+} V_- V_+
\nonumber\\ &&
  +2 F_{2-} \mu V_++2 \mu F_{IR} V_-,
\nonumber\\[0.2cm]
  2\theta_{33}&=&4 m_h^2 F+2 m_h^2 F_{d} \tau^2-F_{d} \mu \tau V_--F_{1+} \mu V_+,
\nonumber\\[0.2cm]
  2\theta_{42}&=&-2 F_{d} S_p \tau V_++F_{1+} S_p V_-+F_{1+} S_x V_+
\nonumber\\ &&
  +2 F_{2-} \mu S_p+2 F_{2-} V_++2 \mu F_{IR} S_x
\nonumber\\ &&
  -4 F_{IR} S_m \tau+2	F_{IR} V_-,
\nonumber\\[0.2cm]
  2\theta_{43}&=&4 F S_m-F_{d} \mu S_x \tau+2 F_{d} S_m \tau^2-F_{d}  \tau V_-
\nonumber\\ &&
  -F_{1+} \mu S_p-F_{1+} V_+.
\nonumber
\end{eqnarray}
Here $V_{\pm}=V_1\pm V_2$,
\begin{eqnarray}
F_{d}&=&\frac{1}{\kappa_1\kappa_2},
\\
F_{1+}&=&\frac{1}{\kappa_1}+\frac{1}{\kappa_2},
\\
F_{2\pm}&=&\biggl(\frac{m^2}{\kappa_2^2}\pm\frac{m^2}{\kappa_1^2}\biggr),
\\
F_{IR}&=&F_{2+}-Q^2F_d.
\end{eqnarray}
where $\kappa_{1,2}$ are defined in Eq. (\ref{kappa}).

\begin{eqnarray}
\theta_{52}&=&
4 \biggl(2 F_{11} E_2 +2 F_{22} E_1 + F_d (E_{12} \tau-Q^2(E_1+E_2))
\nonumber \\ &&
+\frac{2}{S}(E_1S - E_{12}\tau + E_{12} - E_2S)\biggr),
\nonumber \\
\theta_{53}&=&
2 \biggl(F_{1+} (E_2-E_1) - F_d (E_1+E_2) \tau
\nonumber \\ &&
+\frac{4}{S}F_{11} E_2
(\tau -1)\biggr),
\end{eqnarray}

\begin{eqnarray}
E_{12}&=&\epsilon(k_1,k_2,p_1,p_h)=
\frac{1}{4}\biggl[
-(V_1
X-Q^2 S_m+S V_2)^2+
\nonumber \\   &&
+4 (SX-M^2 Q^2) (Q^2 m_h^2-V_1V_2)\biggr]^{1/2},
\nonumber \\
E_{1}&=&\epsilon(k,k_1,p_1,p_h)=
\frac{1}{4}\biggl[
- (R_w V_1 (\tau-1)-R_w \mu S+S_m \kappa_1)^2+
\nonumber \\   &&
+4 (\kappa_1 M^2+R_w (\tau-1) S) (-\kappa_1 m_h^2+V_1R_w\mu)\biggr]^{1/2},
\nonumber \\
E_{2}&=&\epsilon(k,k_2,p_1,p_h)=
\frac{1}{4}\biggl[
- (R_w V_2 (\tau-1)-R_w \mu X+S_m \kappa_2)^2+
\nonumber \\   &&
+4 (\kappa_2 M^2+R_w (\tau-1) X) (-\kappa_2 m_h^2+V_2R_w\mu)\biggr]^{1/2}.
\end{eqnarray}

%
%
%

\end{document}